# Picometer-resolution dual-comb spectroscopy with a free-running fibre laser

Xin Zhao[1], Guoqing Hu[1], Bofeng Zhao[1], Cui Li[1], Yingling Pan[1], Ya Liu[1,2], Takeshi Yasui[3-4] and Zheng Zheng[1,2]*

[1] School of Electronic and Information Engineering, Beihang University, 37 Xueyuan Rd, Beijing 100191, China
[2] Collaborative Innovation Center of Geospatial Technology, 129 Luoyu Road, Wuhan 430079, China
[3] Institute of Technology and Science, Tokushima University, 2-1 Minami-Josanjima, Tokushima 770-8506, Japan
[4] JST, ERATO, MINOSHIMA Intelligent Optical Synthesizer Project, 2-1 Minami-Josanjima, Tokushima 770-8506, Japan

*zhengzheng@buaa.edu.cn

**Abstract:** Dual-comb spectroscopy utilizes two sets of comb lines with slightly different comb-tooth-spacings, and optical spectral information is acquired by measuring the radio-frequency beat notes between the sets of comb lines. It holds the promise as a real-time, high-resolution analytical spectroscopy tool for a range of applications. However, the stringent requirement on the coherence between comb lines from two separate lasers and the sophisticated control system to achieve that have confined the technology to the top metrology laboratories. By replacing control electronics with an all-optical dual-comb lasing scheme, a simplified dual-comb spectroscopy scheme is demonstrated using just one dual-wavelength, passively mode-locked fiber laser. Dual-comb pulses with a repetition-frequency difference determined by the intracavity dispersion are shown to be sufficiently stable against common-mode cavity drifts and noises. As sufficiently low relative linewidth is maintained between two sets of comb lines, capability to resolve RF beat notes between comb teeth and picometer-wide optical spectral features is demonstrated using a simple data acquisition and processing system in an all-fiber setup. Possibility to use energy-efficient, free-running fiber lasers with a small comb-tooth-spacing could enable the realization of low-cost dual-comb spectroscopy systems affordable to more applications.

# Picometer-resolution dual-comb spectroscopy with a free-running fibre laser

## 1. INTRODUCTION

With a large number of equally-spaced, sharp spectral lines covering a wide spectral range, optical frequency combs [1,2] has become a precise metrology tool that enables many high-sensitivity sensing applications previously performed with continuous-wave lasers. Among such applications, the development of laser comb technologies have enabled a number of comb-based spectroscopy schemes covering the optical, infrared, ultraviolet, and even terahertz wavelength windows [3-6]. Among them, dual-comb spectroscopy [7-12] offers the prospect of surpassing the widely used Fourier transform spectroscopy (FTS) techniques [13,14]. In a conventional dual-comb system, two mode-locked ultrafast lasers with their lasing mode spacing and carrier-envelop offset frequencies stabilized are used. Their output pulse repetition rates $f_{rep}$ are meticulously synchronized to be almost the same but with a small frequency difference $\Delta f$. This frequency difference enables real-time temporal pulse walk-off between the pairs of pulses from the two lasers and, thus, asynchronous cross-sampling between them [15-17]. The Fourier transform of the temporal asynchronous optical sampling (ASOPS) cross-correlation waveform, which is analogous to the traditional FTS, reveals the spectral beatings between the pairs of adjacent comb teeth. Thus, the spectral information from the sample such as absorption at certain wavelengths can be obtained. Because of the attainable large stretch factor, broad optical spectrum could be potentially mapped onto a much narrower radio frequency counterpart. The spectral measurement resolution could reach the comb tooth spacing, which can be on the order of MHz for fiber lasers. Molecular study, remote sensing, and environmental monitoring are potential candidates for this attractive technology's applications. Additionally, emerging photonic technologies, such as novel advanced nanophotonic devices with unprecedented cavity finesse, now can possess complicated and sharp spectral features in the near-infrared region [18-20]. Fast spectral measurement with picometer or better spectral resolution is demanded by many such applications.

While dual-comb spectroscopy holds promise to dramatically increase the measuring speed and resolution over FTS, the challenges of developing the laser sources used in these systems remain huge. As the two pulse sequences are generated by two mode-locked lasers with two independent cavities, independent random cavity drifts would quickly destroy the coherence between the comb lines from these lasers. Thus, accurate cavity controls and exquisite care are needed to control the repetition rates and the carrier-frequency offset of the two lasers. Yet, even for two comb sources that are individually-stabilized to radio frequency (RF) references, whose control system would compensate for long-term variations, the short-term stability required by dual-comb spectroscopy may not be sufficiently ensured [11]. One needs to further synchronize the pulses from two independent lasers with good enough precision for interferometric measurements. Therefore, despite the tremendous improvements in the mode-locked lasers, such stabilized dual-laser comb systems still remain costly, complicated and bulky, due to the complex control subsystems required. Therefore, in order to open up new opportunities such as field-based remote-sensing [21] and other real-world applications beyond laboratory studies, the dual-comb technique not relying on the state-of-the art comb systems has been an area that attracts much research interests.

Different schemes to alleviate the complexity of dual-comb systems have been investigated. One kind of such approaches had resorted to the deployment of advanced data acquisition systems to adaptively adjust the sampling clocks in real-time [11, 23]. Instead of electronically stabilizing the repetition frequencies, the random drifts of the two free-running mode-locked lasers need to be electronically tracked and post-compensated. Compared to its widely-adopted FTS predecessors based on a simple optical interferometric setup, the dual-laser source and complicated and high-performance electronic control or data acquisition system still pose a challenge for further development and wider adoption of the dual-comb spectroscopy techniques. Breakthroughs in a drastically simplified system structure are desired to truly surmount these technical barriers. Frequency comb sources based on modulating a cw laser with electro-optic modulators (EOM) is another option besides mode-locked lasers. Their repetition rate can be flexibly set and the comb teeth have good mutual coherence [24]. By spectrally broadening the combs in nonlinear fiber, ~300 GHz bandwidth EOM-based dual-comb spectroscopy has been recently demonstrated [25].

On the other hand, rather than relying on the electronics, exploring all-optical schemes to generate the needed dual-comb pulses is another alternative for low-complexity dual-comb metrology techniques. It had been proposed and demonstrated in the past few years that simultaneous, mode-locked lasing in a single cavity could be enabled through different cavity multiplexing schemes in fiber lasers [26-29]. Some interesting comparison could be drawn to those wildly successful fiber communication multiplexing technologies, which leverage multiplexing in different physical dimensions to greatly expand the communication bandwidth. In such 'multiplexed' mode-locked fiber lasers, the difference in the repetition rate could be set by the intracavity group velocity dispersion [27] as well as birefringence [26, 28, 29].

Asynchronous optical sampling measurement using a dual-wavelength, dual-comb fiber laser was demonstrated for the pump-probe measurement [30] and later for ranging as well [31], showing the potential of the single-laser scheme as a compact and simple alternative for dual-comb systems. More recently, inspired by the potential dual-comb applications, dual-comb lasing had been realized in more mode-locked laser platforms such as solid-state [32, 33] and semiconductor disk lasers [34] with gigahertz pulse repetition rates.

Here we demonstrate a dual-comb spectroscopy scheme based on a free-running, dual-wavelength fiber ring laser that achieves picometer spectral resolution using the simple constant-clock sampling and data processing process. Due to their relatively long cavity lengths, fiber lasers with pulse repetition rates on the order of tens of MHz could be suitable for high-resolution spectroscopy measurement. Compact, power efficient, and cost-effective all-fiber setup could enable the development of robust and affordable benchtop dual-comb spectroscopy instrument.

## 2. METHODS

Our setup shown in Fig. 1. is a typical basic dual-comb spectroscopy one but replacing the dual-laser comb system with a dual-wavelength mode-locked fiber laser. The unidirectional-oscillating fiber laser generates two ultrashort pulse trains centered near 1533 and 1544 nm, respectively, with similar but slightly different repetition rates around 52.74 MHz. Two trains of ultrashort pulses are separated by a bandpass filter into two arms of a fiber interferometer. In each arm, the pulses are amplified respectively by an Erbium-doped fiber (EDF) amplifier constructed with EDF of large normal dispersion, and the pulses experience strong chirp as well as large gain through the gain fiber. They are spectrally-broadened in the EDF and the ensuing standard single-mode fiber. A programmable spectral filter is used to control the bandwidth of the 1544 nm pulse incident onto the detector, in order to avoid frequency aliasing [10]. The 1533 nm pulse passes through the sample under test, which is either a microring resonator or a 10-cm-long gas cell containing 20-torr Acetylene $12C_2H_2$. The reference optical path is not used when the sample under test is the microring resonator to avoid the influence on the longer tail of the interferogram signal resulted from the sharper spectral resonance dip. When the sample under test is the gas cell, the reference optical path consisting of 0.6 m-long single mode fiber is used. Polarization controllers are used to align the state of polarization for coherent detection using by a balanced photodetector. Attenuation is also applied to avoid unwanted nonlinear effects in the rest of the setup and saturation of the photodetector. The 1533 and 1544 nm pulse trains are re-combined by a 3-dB fiber coupler and generate the ASOPS interferogram at an InGaAs balanced photodiode (Thorlabs PDB420). Except the coupling to and from the sample, the setup is all-fiber with more than 20 m of fiber in each arm under an unstablized environment. The time domain electrical interference signal is digitized with a 100M Sample/s, 14bit data acquisition board (NI-5122) with a 16M Sample on-board memory, and the temporal data is further processed by software to provide the spectral information.

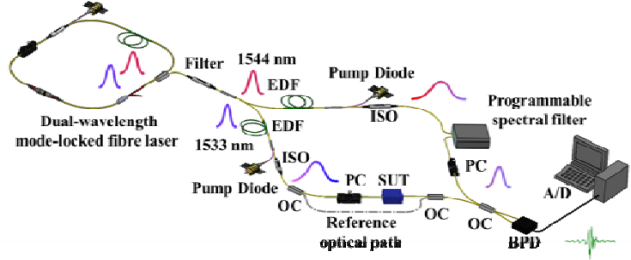

Fig. 1. Dual-comb spectroscopy experimental setup. EDF: Erbium-doped fiber; ISO: isolator; OC: optical coupler; PC: polarization controller; SUT: sample under test; BPD: balanced photodetector.

The EDF-based, dual-wavelength fiber ring laser is mode-locked by a single-wall carbon nanotube (SWNT) modelocker, shown in Fig. 2. The mode-locker is fabricated using the method of optical deposition on an FC/APC ferrule. The laser cavity consists of ~0.43 m EDF (Er110), 0.3-m Hi1060 fiber and ~3.17 m SMF. Besides a polarization-independent optical isolator and a polarization controller (PC), an in-line PBS with ~0.25 m-long polarization maintaining fiber (PMF) pigtails is also placed in the cavity. A hybrid 980/1550nm wavelength division multiplexer/isolator (WDM/ISO) ensures the unidirectional oscillation of the pulses in the cavity. Based on the fiber lengths, the average dispersion is estimated to be anomalous at ~40.2 fs/nm, and the laser is thus operated in the soliton regime. The intracavity birefringence and polarization-dependent loss result in spectral filtering [35] and intracavity loss tuning [27] effects, which is leveraged to control the dual lasing wavelengths of the laser.

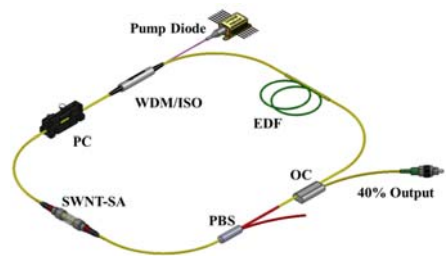

Fig. 2. Dual-wavelength mode-locked fiber laser.

A SiN microring device is used in our current experiments as a test sample, which has a radius of 100 μm and a cross section of 2 μm by 500 nm. A straight waveguide with the same height and a width of 1 μm is coupled to the microring. The gap between them is 700 nm wide. The device is coupled using lensed fiber with a fiber-to-fiber insertion loss of 5 dB for the TM polarization.

## 3. RESULTS

When the laser is pumped above the mode-locking threshold and the intracavity cavity PC properly adjusted, two spectral peaks with similar intensity can emerge in the output spectrum of the laser (shown in Fig. 3(a)). Their 3-dB spectral bandwidths are 2.7 nm and 3.3 nm, respectively. The repetition rates and their difference measured by a fast photodetector (New Focus 1611) and an RF spectrum analyzer (Agilent E4404B) are 52.743118 MHz, 52.744368 MHz and 1250 Hz, respectively, as seen in Fig. 3(b). The value of $\Delta f$ is consistent with the total group velocity dispersion in the fiber cavity and the 11-nm center wavelength separation, rendering it a parameter that can be varied by changing either the intracavity dispersion or the center wavelength separation. The non-aliasing bandwidth for dual-comb spectroscopy is estimated to be 8.8 nm. The 1544 nm pulse is amplified from 0.14 mW to 25 mW, and the 1533 nm one to ~16 mW. The latter is amplified to a lower power level since it needs less spectral broadening as the spectral range of interest in this demonstration is mostly between 1530 and 1540 nm. After amplification and spectral broadening, the spectral widths at -10 dB from their peaks of the pulses become 33 nm and 22 nm, respectively. A 16–nm-wide window with sufficient spectral overlap becomes available to cover the 1528 to 1543 nm spectral range of interests in this demonstration (see Fig. 3(a)).

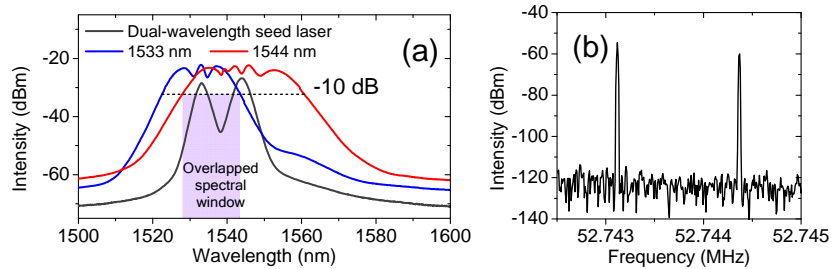

Fig. 3. (a) Output optical spectra of the dual-wavelength seed laser and that after amplification and spectral broadening; (b) RF spectrum of the dual-comb output of the dual-wavelength laser.

To more accurately monitor the repetition rates and their difference of the dual-wavelength pulses whose stability is critical for the dual-comb applications, they are further experimentally measured. The repetition rates are monitored by measuring the down-converted, low-frequency beat signal between the pulse trains and a ~6 GHz sinusoidal signal from a RF synthesizer. The RF frequency is close to the ~113th order of $f_{rep}$, and by monitoring the high order harmonic of the repetition rates, higher measurement accuracy can be achieved. As expected from a free-running fiber laser without active cavity control, due to the environmental instability like temperature changes, both repetition rates drift. Yet, as shown in Fig. 4, their difference $\Delta f$ remains very stable, as both repetition rates are drifting in the same direction and by the same amount. The standard deviation of $\Delta f$ is 16 mHz, while both repetition rates change by more than 10 Hz.

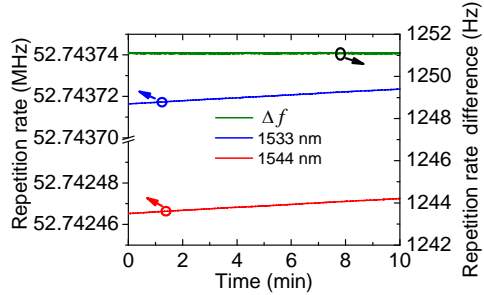

Fig. 4. Monitored variations in the repetition rates of the dual-wavelength mode-locked fiber laser output and their difference.

The relative linewidth of dual-comb source is another important factor that may limit the spectral resolving capability in a dual-comb system. If the relative linewidth is to be much larger than $\Delta f$, the spectral resolution of the dual-comb spectroscopy system would be severely compromised [36]. Here the relative linewidth of the dual-comb lines after spectral broadening is measured using a narrow-linewidth CW laser at 1549.93 nm wavelength (NTK Photonics E15, linewidth < 0.1 kHz). The laser's output beats with the pulse trains after being filtered separately with 0.3 nm-bandwidth bandpass filters, respectively, at two low-pass photodetectors, and their outputs are further mixed. The signal is digitized by the data acquisition board, and the result is Fourier transformed to obtain its spectrum. Since this wavelength, determined by the operating wavelength of the CW laser, is beyond the overlapped spectral window in this demonstration, the signal from the 1533nm pulse is broadened with higher pump power to get enough signal power in the filter band. As shown in Fig. 5, the width of the sharp peak in the Fourier transformed curve of 20-ms long sampled temporal signal is ~250 Hz. This value is much less than $\Delta f$. Further improving the laser, amplifier and spectral broadening stage [37] could improve the linewidth and reduce this part of the measurement errors.

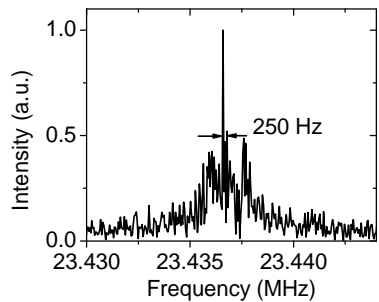

Fig. 5. Relative linewidth measurement between two comb lines from the two pulse trains after spectral broadening.

The ASOPS interferogram generated by the optical field interference between the dual-comb pulses is shown in Fig. 6(a), which is sampled at a constant 100 MS/s rate when the power before the BPD is ~3 μW. To illustrate the spectral measurement resolution of our scheme, the transmission spectral characteristics of the high-Q on-chip microring resonator device is first tested. Because of the large waveguide cross-section, multiple transverse electric (TE) and transverse magnetic (TM) modes are supported. Within one free spectral range (FSR) of ~1.8 nm, four sharp spectral dips are present, corresponding to two TE and two TM modes of the ring. Their quality factors vary from $3*10^5$ to $10^6$ around 1540 nm. Between 1541 nm and 1543 nm, the resonance modes have spectral dips with FWHM widths ranging from 5 pm to around 1 pm. There are also a few much shallower and wider dips caused by a similar microring yet non-optimally coupled to the same bus waveguide. After the signal light goes through the sample, the temporal trace shows a longer and small tail (see Fig. 6(a)). A portion of the RF power spectrum generated from Fourier transform of 0.16 sec of data after applying the linear phase correction [38, 40] is shown in Fig. 6(b). It shows clearly resolved beatnotes as well as a zoom-in subplot showing the noise at other frequencies. The direct observation of beatnotes between each pairs of frequency-comb modes illustrates the good, inherent coherence between the dual-comb pulses. It shows the potential of free-running dual-comb fiber lasers for high-resolution, comb-tooth-resolved dual-comb measurements, in contrast to traditional free-running, dual-laser schemes [12]. The optical spectra obtained by averaging over 5 or 199 interferograms, respectively, in the Fourier domain after phase correction in software [36, 38] are shown in Fig. 6(c). Though the averaged spectrum of 5 interferograms is noisy and the spectral features are not easy to identify, after averaging over 0.16 second of data, which include 199 interferograms, sharp dips can be resolved. The sharp spectral dips occur at the spectral positions when light is coupled into the corresponding high-Q modes of the device. The normalized rms $\sigma_H$ is defined as the standard deviation of the spectral curve normalized by its corresponding average magnitude. The inset shows the inverse of $\sigma_H$, i.e. the spectral signal to noise ratio (SNR) [39], improves from 7 to 113 when the number of averaged interferograms increases from 1 to 199. The corresponding figure of merit of SNR, the quality factor, is calculated as the product of the SNR per unit acquisition time and the number of spectral elements [39], and is $\sim 1*10^6$ in this case.

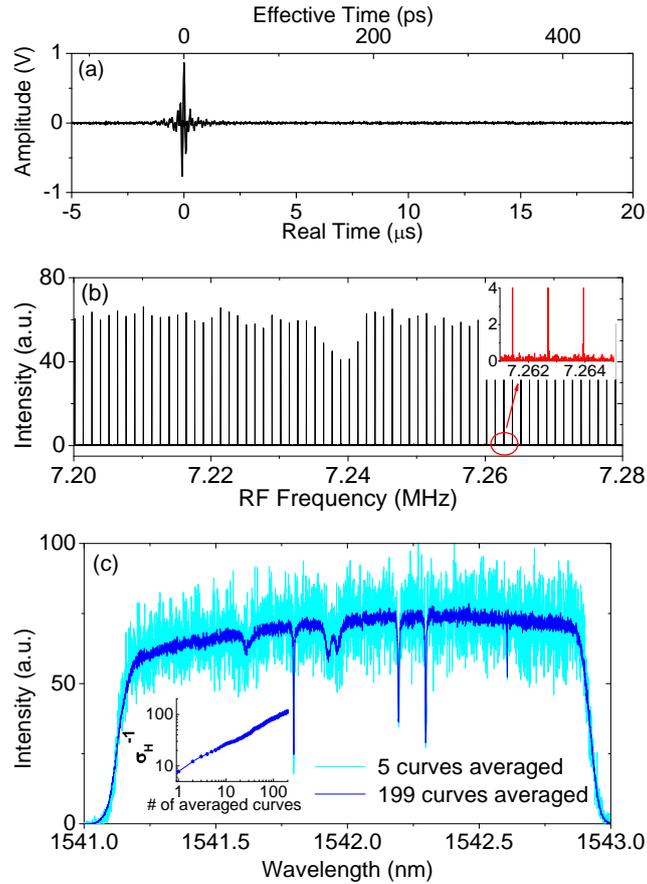

Fig. 6. ASOPS temporal interferogram and its Fourier transformed spectrum. (a) ASOPS temporal interferogram when the 1533 nm amplified pulse passes through the microring resonator device. (b) A comb-tooth-resolved expanded view near a transmission dip of the Fourier transformed RF spectrum of ~0.16 s data. The inset is a zoom-in near the bottom of the comb teeth. (c) Fourier transformed spectrum with different averaging lengths. The programmable spectral filter's passband is set between 1541.1 nm and 1543 nm, which determines the spectral measurement range.

As the rather small on-board memory of our data acquisition board limits our continuous acquisition time to 0.16 second under our sampling rate, to get longer data sequences for further increasing the SNR, a simple approach is taken: The interferograms are continuously acquired until the on-board memory is full; then the data are transferred to a computer to free-up the memory; the next round of acquisition can re-start, until the needed length of data are gathered. This approach can accumulate a relatively long data sequence, without using the advanced DSP board or on-board signal processing. Unlike many previous demonstration using advanced data acquisition systems, where the data are continuously acquired or simultaneously processed, our current low-complexity solution leaves 'gaps' between the acquired pieces of data, as it takes a couple of seconds to upload the data from the board to PC when the measurement is stopped.

Therefore, the experimental time required would be significantly longer than the acquisition time to obtain that length of data. It demands that system, especially the dual-comb source, remains stable enough for the spectral averaging over a longer than normal period of time. To justify the validity of our approach and verify the long-term and short-term stability of our dual-wavelength laser, we compare the results obtained from a continuously acquired data and another set measured by intentionally 'slowing down' the process by 1000 times. As illustrated in Fig. 7, by making 16 separate, shorter measurement spanning over a period of 160 s under the 'slow down' mode, the result shows little degradation when compared with the one measured continuously in 160 ms. Both curves have almost the same shape including the sharpest feature shown in the inset. The spectral position of the dip in the inset also corresponds to that shown in Fig. 6(b). This shows that even over a relatively long period of time, the drifts in the laser comb lines don't significantly deteriorate the quality of data for spectral averaging.

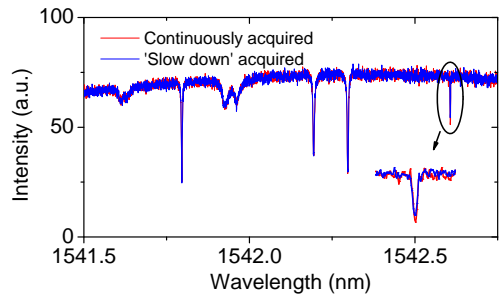

Fig. 7. Comparison of measured spectra after passing through the microring resonator device under different acquisition modes. Red: spectrum averaged from ~192 interferograms continuously acquired in less than 0.16 second; Blue: one from the same number of interferograms when the data are gather sequentially in 16 'slices', each containing 12 interferograms, over a period of 160 seconds. Inset: zoom-in of the sharpest dip.

By acquiring 1990 interferograms based on the above data acquisition scheme, the transmission of the microring resonator is measured as shown in Fig. 8. The result is compared with that obtained by scanning a narrow-linewidth CW tunable laser (Agilent 8610B, 0.1 pm wavelength resolution) at a 0.5 nm/min speed and measuring the transmission power with a power meter. The state of polarization of the laser is set to excite both the TE and TM modes simultaneously. The sharpest dip among the multiple resonances within one FSR has a FWHM width of 1.2 pm as measured by the scanning laser method. The FWHM width is measured to be 1.5 pm in the dual-comb spectroscopy result, which matches the laser scanning data relatively well. Compared to that measured by scanning a high-end tunable laser and, our result clearly resolves the sharpest dip about slightly wider than 1 pm wide, just a few times of the comb tooth spacing. This indicates that the spectral resolving capability of our low-cost setup has the potential to approach that of many stabilized dual-comb systems [12]. We note that the shape of the dip in the laser scanning result is distorted and deviates from the more symmetric intrinsic spectral lineshape due to the thermal loading effect after the injection of the CW light into the resonator, while our dual-comb result gives a symmetric yet slightly wider shape. It is also noted that because of the difference in the coupling conditions to the devices and different excitation state of polarization that results in varied coupling into different modes, the depths of the spectral dips could be somewhat different between our measurement and the scanning one.

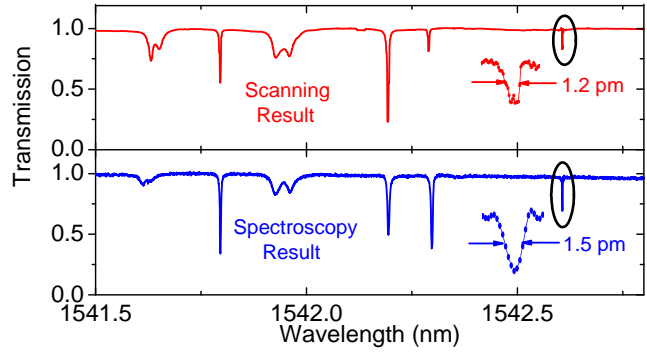

Fig. 8. Measured transmission spectrum of the microring resonator device with transmission dips as narrow as 1.2 pm.

To demonstrate the broadband sensing capability and also quantify the resolution and accuracy of our scheme, the absorption spectrum of a gas cell containing acetylene is measured between 1528 and 1543 nm, whose spectral features are well-defined and documented. The short wavelength limit is limited by the filtering range of the programmable filter used. To measure the spectral transmission, the reference path is used in parallel to the gas cell to generate a reference pulse undisturbed by the gas. The relative delay between the signal and reference pulse is ~9.3 ns, approximately half a period between pulses, so that its interferogram doesn't affect that of the signal. In this self-referenced architecture, one period of the interferogram contains both a signal and a reference, and both spectra can be retrieved, so can the transmission. Our current spectral measurement range is set by the tuning range of the programmable filter. To avoid aliasing, the transmission spectrum is measured over a 2 nm window each time and then spectral stitching is applied to obtain the transmission spectrum. 1990 interferograms are averaged to yield each piece of the spectrum, corresponding to acquiring data over a period of 1.6 s. Besides the strong rovibrational lines of the $v_1 + v_3$ band of acetylene, the weak lines from its hot bands are also well-preserved. Both these strong lines and weak lines of the experimental result match the calculated HITRAN results based on the gas parameter, both in its amplitude and spectral shapes, as shown in Fig. 9(a). This also illustrates that the spectrally broadened spectra of pulses with different center wavelength can maintain their spectral coherence at a level suitable for spectroscopy measurements. The FWHM width of P19 line and P20 line are measured to be 7.0 pm and 5.1 pm in the experimental result while the theoretical value is 6.1 pm and 4.9 pm.

Good agreement between the experimental spectrum with the HITRAN data confirms the potential of our technique for high-resolution spectral measurements. The capability of averaging over many interferograms without causing significant distortion of the observed spectrum under a relatively low SNR show the stability of the dual-comb laser against common-mode noise. In contrast, for dual-laser schemes, the directly averaged spectra under a constant clock would be complete smeared due to the random walk between the two uncorrelated lasers and do not allow for averaging [11]. Our results demonstrate that, despite the relatively narrow initial spectral widths, the dual-comb laser we propose to use here is capable of being applied to a wider spectral range through relatively simple spectral broadening.

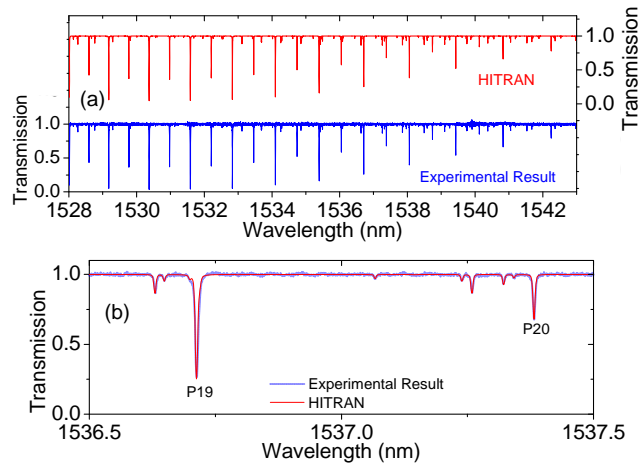

Fig. 9. Experimental transmission spectrum of the $12C_2H_2$ gas cell. (a) Experimental transmission spectrum of $12C_2H_2$ sampled at the constant clock of the DAC board, which is compared with a spectrum calculated by HITRAN. (b) The zoom-in spectrum of a span of 1 nm around the P19 line and P20 line of $12C_2H_2$.

## 4. DISCUSSION

It should be noted that, compared to the highly-stabilized dual-comb systems, non-negligible spectral measurement uncertainty in our scheme remains, as shown above. This is resulted from the nature of the free-running laser that unavoidably brings more jitter and drift into the comb lines, a price to pay for a much simplified system. Besides the relative linewidth of the dual-comb pulses, by using a free-running fiber laser source, there are some additional factors that may affect the achievable spectral resolution. As in this demonstration our laser cavity is not actively environmentally controlled, the repetition rates would slowly drift as affected by temperature and other factors. With a change of 1 Hz in the repetition rate, in the optical spectral window we are investigating, the comb lines would shift ~3.7 MHz (~0.03pm). If the data are acquired in a couple of minutes when the repetition rate drifts a couple of Hz per minutes, it would cause a few tenth of picometer wavelength uncertainty, even when $\varDelta f$ remains very stable. Laser packaging and cavity stabilization schemes that are common for many commercial lasers can be implemented to reduce such uncertainties.

Also, it is noted that unlike the comb systems where the position of each comb line could be traced and locked to a specific frequency, it is not possible to do that for our system. Yet, the prospect of locking one of the dual-comb outputs to a known standard to be a possible remedy for this shortcoming.

For measuring a broadband spectral response, it would be necessary for the fiber laser output to be nonlinearly broadened to cover a larger spectral range that comb spectroscopy claims to be superior to previous schemes. Our current results indicates that, even after nonlinear spectral broadening, the spectral coherence and comb line correlations between the dual-comb signal generated by the dual-wavelength laser could be maintained.

## 5. CONCLUSIONS

We demonstrate the feasibility of constructing the so-far highly envied yet complex dual-comb spectroscopy system with three simple pieces built with widely available, low-cost commercial parts and modules: one free-running fiber laser, simple standard data acquisition electronics, and commonly used post-processing algorithms. The correlation between two pulse trains oscillating in the same laser cavity enables natural locking of their comb characteristics. In contrast to the approaches using two lasers, two mode-locked lasers' would randomly walk away from each other, our simple dual-wavelength fiber laser that generates well-determined $\varDelta f$ by intracavity group velocity dispersion also shows surprising robustness to maintain frequency correlations between the two pulse sequences. Our results and analyses unequivocally show the capability of such single-fiber-laser-based dual-comb spectroscopy system to reach a spectral resolution good enough for many real-world applications.

Our proposed single-fiber-laser-based dual-comb system could result in significant simplification in the implementation complexity and costs of dual-comb spectrometers. Low-cost and portable ones built with standard off-the-shelf fiber-optic components would be more accessible and affordable to many more applications.

The work at Beihang University was supported by the 973 Program (2012CB315601); NSFC (61221061/61107057/61435002); The Fundamental Research Funds for the Central Universities.

We thank Prof. Andrew M. Weiner and Minghao Qi from Purdue University for providing us the integrated device sample, and Prof. Kun Xu and Yitang Dai from BUPT for lending us the narrow-line-width continuous-wave laser.